\documentclass{article}
\usepackage{spconf,amsmath,graphicx,booktabs,multirow}
\usepackage{float}       
\usepackage{tabularx}
\usepackage{enumitem}
\usepackage{dblfloatfix}
\usepackage{hyperref}

\setlist[itemize]{topsep=3pt,itemsep=1pt,parsep=0pt,partopsep=0pt,leftmargin=1.6em}

\renewcommand{\arraystretch}{1.00}
\newcommand{\tfrecall}[2]{#1\,/\,#2}

\makeatletter
\def\proncft@tabletype{table}
\long\def\@makecaption#1#2{%
  \ifx\@captype\proncft@tabletype
    \vskip 2pt
  \else
    \vskip 4pt
  \fi
  \setbox\@tempboxa\hbox{#1. #2}%
  \ifdim\wd\@tempboxa>\hsize
    #1. #2\par
  \else
    \hbox to\hsize{\hfil\box\@tempboxa\hfil}%
  \fi
  \ifx\@captype\proncft@tabletype
    \vskip 4pt 
  \fi
}
\renewcommand{\section}{\@startsection{section}{1}{\z@}%
  {-9pt}{4pt}{\normalfont\normalsize\bfseries}}
\renewcommand{\subsection}{\@startsection{subsection}{2}{\z@}%
  {-6pt}{3pt}{\normalfont\normalsize\bfseries}}
\makeatother

\makeatletter
\newcount\proncft@lasttable
\newcount\proncft@nexttable
\newif\ifproncft@tablebusy
\newcommand{\proncft@trytables}{%
  \ifproncft@tablebusy\else
    \ifinner\else
      \ifnum\interlinepenalty<\@M
        \ifnum\@listdepth=\z@
          \ifnum\proncft@nexttable<\numexpr\proncft@lasttable+1\relax
            \ifdim\pagetotal>\pagegoal \penalty\z@ \fi
            \@tempdima=\pagegoal
            \ifdim\@tempdima=\maxdimen
              \@tempdima=\@colroom
            \else
              \advance\@tempdima-\pagetotal
              \advance\@tempdima-\pagedepth
            \fi
            \@tempdimb=\ht\csname proncft@tablebox@\the\proncft@nexttable\endcsname
            \advance\@tempdimb\dp\csname proncft@tablebox@\the\proncft@nexttable\endcsname
            \advance\@tempdimb 1pt
            \ifdim\@tempdimb<\@tempdima
              \proncft@tablebusytrue
              \nointerlineskip
              \expandafter\box\csname proncft@tablebox@\the\proncft@nexttable\endcsname
              \global\advance\proncft@nexttable\@ne
              \proncft@tablebusyfalse
              \proncft@trytables
            \fi
          \fi
        \fi
      \fi
    \fi
  \fi
}
\NewDocumentEnvironment{flowtable}{+b}{%
  \par
  \global\advance\proncft@lasttable\@ne
  \expandafter\newbox\csname proncft@tablebox@\the\proncft@lasttable\endcsname
  \proncft@tablebusytrue
  \expandafter\global\expandafter\setbox
    \csname proncft@tablebox@\the\proncft@lasttable\endcsname=\vbox{%
      \hsize=\columnwidth \linewidth=\columnwidth
      \begin{table}[H]#1\end{table}%
    }%
  \proncft@tablebusyfalse
  \proncft@trytables
}{}
\AddToHook{para/after}{\proncft@trytables}
\newcommand{\FlushFlowTables}{%
  \par\proncft@trytables
  \ifnum\proncft@nexttable<\numexpr\proncft@lasttable+1\relax
    \newpage\FlushFlowTables
  \fi
}
\makeatother

\newcommand{\etal}{\emph{et al.}}

\title{PronCFT: Pronunciation Contextual Fine-Tuning for
Contextual Biasing of Out-of-Vocabulary Words in LLM-Based ASR}
\name{\fontsize{10.95}{13}\selectfont Yifan Xu$^{1,2}$, Kaixun Huang$^{2}$, Mengzhe Geng$^{1}$, Mingyu Cui$^{1}$, Shen Huang$^{2}$, Shidong Shang$^{2}$, Xunying Liu$^{1}$}
\address{$^{1}$Department of Systems Engineering and Engineering Management,\\
The Chinese University of Hong Kong, Hong Kong SAR, China\\
$^{2}$Tenlai Team, Beijing, China\\
yfxu1203@link.cuhk.edu.hk, xyliu@se.cuhk.edu.hk}

\begin{document}
\ninept
\renewcommand{\baselinestretch}{0.93}\normalsize
\setlength{\abovedisplayskip}{6pt plus 1pt minus 1pt}
\setlength{\belowdisplayskip}{6pt plus 1pt minus 1pt}
\setlength{\abovedisplayshortskip}{2pt plus 1pt}
\setlength{\belowdisplayshortskip}{4pt plus 1pt minus 1pt}
\begingroup
\renewcommand{\arraystretch}{0.72} 
\maketitle
\endgroup
\raggedbottom

\begin{figure*}[!t]
\centering
\IfFileExists{cotoov_v3.pdf}{\includegraphics[width=\textwidth,height=7.2cm,keepaspectratio]{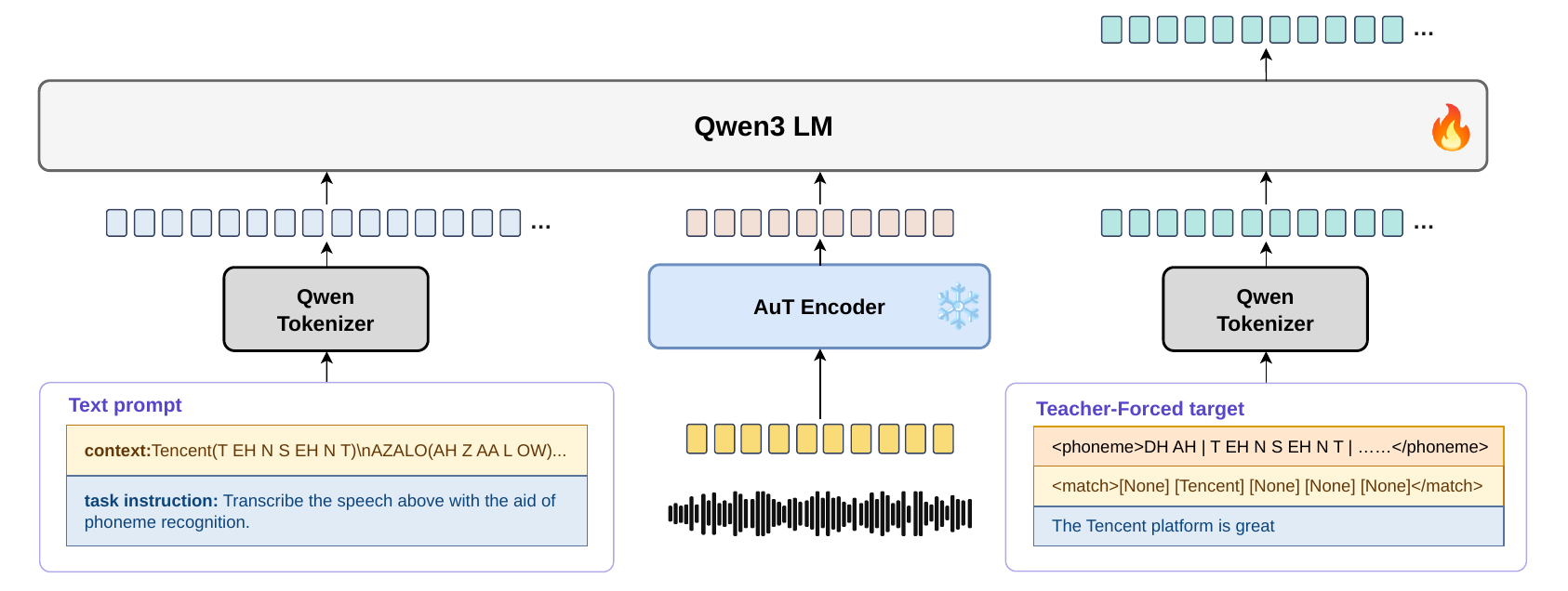}}{\fbox{\parbox[c][7.2cm][c]{0.96\textwidth}{\centering \texttt{cotoov\_v3.pdf} not found.}}}
\caption{Overview of PronCFT. Fixed candidate pronunciations enter
the prompt (1). Structured supervision predicts utterance phonemes
and matches them to candidate pronunciations (2), then outputs the
matched spelling (3). The audio encoder and projector are frozen
while the LLM is fine-tuned.}
\label{fig:pipeline}
\end{figure*}

\begin{abstract}
LLM-based ASR systems often replace rare and out-of-vocabulary (OOV) words with familiar words that sound similar. Two factors contribute to this error: the model has not learned how an unseen word's sound maps to its spelling, and it may favor a familiar word over the one spoken, a failure we call \emph{prior suppression}. Providing candidate pronunciations alone may not help: the model can treat a phoneme annotation as a marker that raises a candidate's output probability without checking its pronunciation against the speech. We propose PronCFT, which uses fixed phoneme sequences and trains the model to transcribe speech into phonemes, match them to context candidates, and output the matched spelling. During training, we alter candidate spellings while keeping their pronunciations unchanged, discouraging the model from choosing a familiar spelling without checking how the candidate sounds. A curriculum then gradually removes intermediate outputs to reduce decoding latency and prevent phoneme transcription errors from affecting the final text. The model's retained ability to use pronunciation cues suggests that this ability can be internalized. On an OOV test set filtered from open-source benchmarks, the final model, PronCFT-C, significantly improves recall from 59.89\% to 83.57\% (+23.68\,pp). Wrong-phoneme and spelling-interference tests indicate that the gain depends on phoneme content rather than the presence of an annotation.
\end{abstract}

\begin{keywords}
Speech recognition, contextual biasing, out-of-vocabulary words, phoneme grounding, large language models
\end{keywords}

\section{Introduction}
\label{sec:intro}
LLM-based ASR systems~\cite{radford2023whisper,bai2024seed,chu2024qwen2audio,shi2026qwen3asr}, built on the encoder--adaptor--LLM paradigm, have achieved strong general-ASR performance, yet they remain vulnerable to rare named entities and out-of-vocabulary (OOV) words, such as names, brands, and domain terms. Mistranscribing these words can change the meaning of a transcript. Contextual biasing supplies candidate words to the model~\cite{pundak2018clas,sun2023tcpgen,liu2025trie}, and our method builds on this approach. We identify two failure modes:
\begin{itemize}
\item \textbf{Prior suppression}: the decoder may struggle to map speech to an unseen spelling or may favor a phonetically similar common word over the intended one (\emph{cloud} instead of \emph{claude}).
\item \textbf{Marker shortcut}: when phoneme annotations are supplied without explicit alignment supervision, their presence can make the model more likely to output an annotated candidate even if its phoneme sequence is incorrect. The model therefore need not match the phoneme content to the speech.
\end{itemize}

These failures persist in current open systems despite strong general WER~\cite{shi2026qwen3asr,xu2025fireredasr,wang2026covo}. Pronunciation-aware biasing methods such as PAC~\cite{fu2025pac} add phoneme annotations to the context, but do not explicitly train the model to align those phonemes with acoustic evidence. We investigate whether fixed phoneme sequences can serve as candidate-level cues whose content, rather than mere presence, is used during supervised fine-tuning. Counterfactual phoneme swaps and corruption show that, in the tested prompt-only setting, phoneme annotations can act as candidate markers even when their content is wrong (Section~\ref{ssec:expinterfere}).

We therefore introduce \textbf{PronCFT} (pronunciation contextual fine-tuning), which trains the LLM to use fixed phoneme sequences as pronunciation evidence. Its structured training target jointly supervises utterance phoneme transcription, matching between the utterance and context candidates, and final transcription. The model thus learns to compare the sounds in speech with candidate pronunciations before outputting a spelling. We also change candidate spellings while preserving their pronunciations during training, discouraging the model from choosing a familiar spelling without checking how the candidate sounds. A curriculum then gradually removes the auxiliary outputs so the model can use what it has learned without generating intermediate steps, reducing decoding latency, and preventing intermediate errors from propagating to the final text.

PronCFT reaches 74.12\% OOV recall, compared with 59.89\% for the basemodel. The final curriculum-trained model, PronCFT-C, further improves recall to 83.57\%. Recall alone cannot distinguish pronunciation-guided decoding from memorized mappings between speech patterns and spellings. We therefore use spelling-interference experiments and attention analysis to examine whether the gains depend on pronunciation evidence.

Our contributions are as follows.
\begin{itemize}
\item Counterfactual evidence that, in the tested prompt-only setting, phoneme annotations can act as candidate markers rather than pronunciation evidence;
\item \textbf{PronCFT}, which jointly supervises phoneme transcription, candidate matching, and final transcription; spelling augmentation discourages reliance on familiar spellings, while curriculum training removes intermediate outputs;
\item An OOV evaluation suite built from open-source benchmarks; inference code and test sets are released at \url{https://github.com/gcsjsdsd/PronCFT.git}.
\end{itemize}

\section{Related Work}
\label{sec:related}

\subsection{Contextual biasing for ASR}
\label{ssec:biasing}
Contextual biasing supplies an ASR system with candidate words to improve recognition of rare entities and out-of-vocabulary (OOV) terms. Existing methods encode candidates in attention~\cite{pundak2018clas}, constrain or reweight decoding~\cite{sun2023tcpgen}, add pronunciations to lexical search~\cite{nakagome2025wctc,liu2025trie}, retrieve relevant candidates~\cite{kong2025glclap}, or train the model to use candidate contexts~\cite{kong2025hotwordrl}. PronCFT follows the training-based approach while pairing each candidate spelling with its phonemes. Recognition gains alone, however, do not show whether a model uses the phoneme content or simply favors an annotated candidate.

\subsection{Phonemes used in E2E and LLM-based ASR}
\label{ssec:pron}
In conventional end-to-end (E2E) ASR, auxiliary phoneme labels improve speech representations during training~\cite{toshniwal2017aux,sanabria2018hmtl}. In LLM-based ASR, phonemes attached to context candidates also serve as inference-time cues: the decoder must relate the speech to a candidate pronunciation and then output its spelling. Without supervision for this matching step, phoneme annotations may function as candidate markers rather than pronunciation evidence.

PAC~\cite{fu2025pac} and PARCO~\cite{he2025parco} introduce phonetic cues into contextual biasing. In PAC's training contexts, targets carry phoneme annotations while distractors are spelling-only, so annotation presence may predict the target without requiring the model to use phoneme content. Novitasari~\etal~\cite{novitasari2026common} instead use acoustic cues from partially similar common words when a suitable G2P system is unavailable. Qu~\etal~\cite{qu2023vocab} adapt the acoustic model with TTS-generated OOV speech. PronCFT annotates every candidate with phonemes, supervises speech-to-pronunciation matching, and uses counterfactual interventions to test whether phoneme content affects recognition (Section~\ref{ssec:expinterfere}).

\section{The PronCFT Method}
\label{sec:method}

\subsection{Phoneme-grounded structured supervision}
\label{ssec:methodft}
We fine-tune the LLM in Qwen3-ASR while freezing its AuT audio encoder
and projector. Training examples are constructed from
GigaSpeech~\cite{chen2021gigaspeech} transcriptions. We build a
1.3M-entry pronunciation dictionary by combining
CMUdict~\cite{cmudict} with out-of-dictionary words whose pronunciations
are annotated by Qwen3-235B-A22B~\cite{yang2025qwen3}. For each word
$y_j$ in an utterance, we retrieve its stress-stripped ARPAbet sequence
$\pi_j$; utterances containing words without pronunciations are discarded.

For each utterance, we select up to eight distinct in-utterance words
and sample 20--50 distractors that are not homophones of any word in
the utterance. We then shuffle the resulting candidate list using a
seed derived from the utterance ID. Every candidate is presented in the system prompt as \texttt{WORD(PHONEMES)}, for example \texttt{claude(K L AO D)}. Thus, the presence of an annotation cannot by itself identify the spoken candidate. We store each candidate's phoneme sequence before any change to its spelling.

The training target has three parts:
\begin{equation}
\underbrace{\pi_1,\ldots,\pi_n}*{\texttt{<phoneme>}}
;\Vert;
\underbrace{m_1,\ldots,m_n}*{\texttt{<match>}}
;\Vert;
y',
\label{eq:target}
\end{equation}
where $m_j$ is the context candidate whose stored phonemes match $\pi_j$, or \texttt{None} if there is no match. The final transcription $y'$ uses a candidate's displayed spelling when that candidate corresponds to a spoken word. Supervising phoneme transcription and candidate matching teaches the model to connect speech, candidate pronunciations, and output spellings. We train the three output parts with teacher forcing and weighted token losses:
\begin{equation}
\mathcal{L}
=\lambda_\pi\mathcal{L}_\pi
+\lambda_m\mathcal{L}_m
+\lambda_y\mathcal{L}_y.
\label{eq:loss}
\end{equation}

\noindent\textbf{Pronunciation-preserving misspellings.}
For selected training examples, we respell a spoken candidate and some distractors while retaining their stored phoneme sequences, for example \emph{tencent}$\to$\emph{tensent}. A replacement is rejected if it occurs in the pronunciation dictionary, with up to ten attempts. The displayed replacement also appears in $y'$. This prevents the model from solving these examples by outputting only the familiar original spelling.

\subsection{Curriculum removal of auxiliary supervision}
\label{ssec:methodcompress}
At L0, the model autoregressively emits phoneme transcription and
candidate-alignment segments before producing the final transcription.
We progressively remove these auxiliary outputs for three reasons.
First, this tests whether the speech--phoneme--spelling associations
learned under structured supervision have been internalized in the model
parameters: a transcription-only decoder should retain the benefit
without explicitly generating the intermediate steps. Second, auxiliary
segments substantially increase the number of autoregressive output
tokens and hence decoding latency. Third, when structured segments are
decoded at inference, an early phoneme or alignment error becomes part of
the autoregressive prefix and can propagate to the final transcription.
Removing the intermediate outputs therefore also eliminates this
error-propagation path.

Starting at 20k updates, we train five stages (L0--L4) for 1.5k updates each, keeping the audio, context prompts, transcription objective, and misspelling augmentation fixed. L1 linearly increases word-level phoneme masking from 0 to 100\%; L2 removes the phoneme segment; L3 retains only matched candidates; and L4 outputs the transcription alone (Table~\ref{tab:compress}).  After L0, auxiliary tokens serve only as transitional outputs: the loss is computed on transcription tokens alone ($\lambda_\pi=\lambda_m=0$ in Eq.~\eqref{eq:loss}).

\begin{flowtable}
\caption{Structured training targets at each curriculum stage for the same utterance, abbreviated with \dots. In L1, \texttt{\_} marks a masked word-level phoneme slot; word-boundary bars are kept.}
\label{tab:compress}
\centering
\footnotesize
\setlength{\tabcolsep}{3pt}
\renewcommand{\arraystretch}{1.08}
\begin{tabularx}{\linewidth}{@{}>{\raggedright\arraybackslash}p{1.35cm}>{\raggedright\arraybackslash}X@{}}
\toprule
Level & Training target (same utterance) \\
\midrule
L0 & \texttt{<phoneme>HH W EH N | K AH N S IH D ER IH NG | \dots\ | S IH CH UW EY T IH D</phoneme><match>[None] [CONSIDERING] \dots\ [None]</match>WHEN CONSIDERING \dots\ SO SITUATED} \\
\addlinespace[3pt]
L1 (40\%) & \texttt{<phoneme>HH W EH N | K AH N S IH D ER IH NG | \_ | \_ | AH V | \dots\ </phoneme><match>[None] [CONSIDERING] \dots\ [None]</match>WHEN \dots} \\
\addlinespace[3pt]
L1 (100\%) & \texttt{<phoneme>\_ | \_ | \dots\ | \_</phoneme><match>[None] [CONSIDERING] \dots\ [None]</match>WHEN \dots} \\
\addlinespace[3pt]
L2 & \texttt{<match>[None] [CONSIDERING] \dots\ [None]</match>WHEN \dots\ SO SITUATED} \\
\addlinespace[3pt]
L3 & \texttt{<match>[CONSIDERING] [ENVELOPE] [WROTE] \dots\ [HAVE]</match>WHEN \dots\ SO SITUATED} \\
\addlinespace[3pt]
L4 & \texttt{WHEN CONSIDERING \dots\ SO SITUATED} \\
\bottomrule
\end{tabularx}
\end{flowtable}

At L1, the masking ratio increases linearly with the iteration,
\begin{equation}
\rho(t) = \min\!\Big(1, \tfrac{t - t_1}{t_2 - t_1}\Big),
\label{eq:mask}
\end{equation}
where $t_1$ and $t_2$ are the start and end iterations of L1. We use
progressive removal rather than abruptly discarding all intermediate
outputs, sequence-level distillation, or final-answer-only RL rewards.
The former may preserve the input-to-transcription mapping without
retaining phoneme-alignment behavior, while the latter do not directly
control contextual shortcuts. By removing one auxiliary component at a
time, the curriculum encourages the model to preserve the learned
associations while transferring them to increasingly efficient
transcription-only decoding.

\FlushFlowTables

\section{Experimental Framework}
\label{sec:exp}

\subsection{Models, data, and metrics}
\label{ssec:setup}

\noindent\textbf{Model and compared systems.}
We use Qwen3-ASR-1.7B~\cite{shi2026qwen3asr}, an
encoder--adaptor--LLM speech recognition model. Its audio encoder
extracts speech features, the adaptor passes them to the LLM, and
the LLM generates the transcription. We keep the acoustic front end
fixed and train the decoder to use candidate pronunciations.
\emph{Giga-SFT}, our training baseline, is fine-tuned without phonemes
or auxiliary targets. \emph{PronCFT} uses structured phoneme
supervision, and \emph{PronCFT-C} is its curriculum-trained version.
We also include reimplementations of PAC~\cite{fu2025pac} and
PARCO~\cite{he2025parco}. Single-output tests use greedy decoding;
compared systems receive the same seeded candidate lists.

\noindent\textbf{Training.}
We train on GigaSpeech~\cite{chen2021gigaspeech} for one epoch,
freezing the audio encoder and adaptor while updating the LLM.
We use bf16 Adam, a global batch size of 384, a maximum sequence
length of 2048, and a WSD learning rate from $10^{-5}$ to $10^{-6}$.
For PronCFT, 50\% of samples include structured auxiliary targets;
65\% of those also use pronunciation-preserving misspellings.
Phonemes are generated by Qwen3-235B-A22B~\cite{yang2025qwen3},
and proper nouns are selected as OOV training anchors.

\noindent\textbf{Evaluation.}
We evaluate general recognition on LibriSpeech test-clean and
test-other~\cite{panayotov2015librispeech}, GigaSpeech
test~\cite{chen2021gigaspeech}, TED-LIUM
3~\cite{hernandez2018tedlium3}, VoxPopuli en
test~\cite{wang2021voxpopuli}, and MLS
test-en~\cite{pratap2020mls}. The merged OOV test set contains
10{,}533 sentences drawn from these benchmarks, VoxPopuli accented,
and CommonVoice en 17.0~\cite{ardila2020commonvoice}. Each sentence
contains a word absent from CMUdict~\cite{cmudict}, our operational
OOV criterion; evaluation OOV words do not overlap with the training
vocabulary. We report word error rate (WER), OOV recall, and spelling
recall under the misspelling and wrong-phoneme interventions.
Following span-wise attention analysis~\cite{clark2019bert}, we
teacher-force the reference transcript and average attention over
layers and heads at target-word positions: $P_t$ and $P_d$ measure
target and distractor phoneme-span attention, and $\mathrm{SAM}_t$
measures target spelling-span attention; we report
$\Delta P=P_t-P_d$ and $\mathrm{PhnR}=P_t/(P_t+\mathrm{SAM}_t)$.
Replacing only the target candidate's phonemes while keeping the
forced output and other candidates fixed gives the paired
control~\cite{hewitt2019control}, with
$\mathrm{PGD}=100(\Delta P^{\mathrm{wph}}-\Delta P)/\Delta P$.
We treat attention as diagnostic rather than causal
evidence~\cite{jain2019attention}.

\subsection{OOV recall on the merged test set}
\label{ssec:exprecall}

\begin{flowtable}
\caption{OOV recall (\%) on the merged CMUdict-filtered test set.}
\label{tab:oov}
\centering
\small
\setlength{\tabcolsep}{4pt}
\begin{tabular}{@{}lrrrr@{}}
\toprule
System & \multicolumn{2}{c}{ASR output}
       & Structured output & 5-best \\
\cmidrule(lr){2-3}
& w/o phn. & w/ phn. & w/ phn. & -- \\
\midrule
Giga-SFT                  & 59.89 & --    & --    & 91.07 \\
PARCO~\cite{he2025parco} & 35.11 & --    & --    & --    \\
PAC~\cite{fu2025pac}     & 59.79 & 63.56 & --    & 95.19 \\
PronCFT                   & \textbf{61.46} & 65.45 & 74.12 & 96.81 \\
PronCFT-C                 & 61.08 & \textbf{70.64}
                          & \textbf{83.57} & \textbf{98.63} \\
\bottomrule
\end{tabular}
\end{flowtable}

We test OOV recall using paired utterances with one spoken target and
20--500 unspoken distractors, with identical seeded candidate lists
across systems. We also report $R@5$ over five independently sampled
transcripts. Without phonemes, PronCFT only marginally exceeds
Giga-SFT (61.46\% vs.\ 59.89\%; not significant). Its recall rises
to 65.45\% with candidate phonemes and 74.12\% when structured output
is added, suggesting that the main gains rely on these cues rather
than a general tendency to emit rare words. Under their reported best
settings, PronCFT exceeds Giga-SFT by 14.23 percentage points
(two-sided exact McNemar test, $p<0.001$). PronCFT-C reaches 83.57\%,
a further 9.45-point gain over PronCFT ($p<0.001$).

\subsection{General ASR performance}
\label{ssec:expgeneral}

Table~\ref{tab:wer} checks whether OOV improvements affect general
recognition. Average WER remains comparable to the Giga-SFT baseline:
6.05\% for Giga-SFT, 6.02\% for PronCFT, and 6.01\% for PronCFT-C.
In-vocabulary WER is also similar (5.72\%, 5.68\%, and 5.75\%,
respectively), while OOV recall improves as shown in
Section~\ref{ssec:exprecall}.

\begin{flowtable}
\caption{General WER (\%) on six open-source test sets.
WER avg. is the macro-average; IV WER scores in-vocabulary
reference words under the criterion of Section~\ref{ssec:setup}.}
\label{tab:wer}
\centering
\small
\setlength{\tabcolsep}{4pt}
\begin{tabular}{@{}lccc@{}}
\toprule
Test set & Giga-SFT & PronCFT & PronCFT-C \\
\midrule
LS test-clean     & 2.23  & 2.22  & 2.20  \\
LS test-other     & 4.53  & 4.56  & 4.61  \\
GigaSpeech test   & 10.60 & 10.59 & 10.56 \\
TED-LIUM 3        & 3.51  & 3.41  & 3.37  \\
VoxPopuli en test & 9.51  & 9.48  & 9.47  \\
MLS test-en       & 5.95  & 5.88  & 5.89  \\
\midrule
WER avg.          & 6.05  & 6.02  & 6.01  \\
IV WER avg.       & 5.72  & 5.68  & 5.75  \\
\bottomrule
\end{tabular}
\end{flowtable}

\subsection{Misspelled-context interference}
\label{ssec:expinterfere}

An OOV recall gain alone does not establish whether a model uses
candidate pronunciations or relies on familiar spellings. We test
this with misspelled candidates and incorrect phonemes. Using the
same OOV utterances and seeded candidate lists as
Section~\ref{ssec:exprecall}, we respell every target and selected
distractors more extensively than during training
(e.g., \emph{tencent}$\to$\emph{tensent}). The new spellings are
absent from the training vocabulary, while their phoneme sequences
remain unchanged. In the wrong-phoneme condition, we replace each
candidate's phonemes with an incorrect sequence of the same length
while keeping its original spelling. Table~\ref{tab:misspelled}
reports original/replacement spelling recall for misspelled
contexts with or without phoneme annotations, and original-spelling
recall for the wrong-phoneme condition. PAC uses the same candidate
lists and respelling procedure.

\begin{flowtable}
\caption{Misspelling and wrong-phoneme interventions (\%).
For misspelled contexts (w/ phn., w/o phn.), cells report
original/replacement spelling recall. For wrong phn. (orig.),
candidate spellings remain unchanged while their phonemes are
corrupted; the value is recall of the spoken OOV word's
original spelling.}
\label{tab:misspelled}
\centering
\small
\setlength{\tabcolsep}{4pt}
\begin{tabular}{@{}lccc@{}}
\toprule
System & w/ phn. & w/o phn. & \shortstack{wrong phn.\\(orig.)} \\
\midrule
Giga-SFT  & \tfrecall{--}{--}
          & \tfrecall{57.62}{0.52} & --    \\
PAC~\cite{fu2025pac}
          & \tfrecall{52.79}{6.11}
          & \tfrecall{51.41}{3.13} & 50.59 \\
PronCFT   & \tfrecall{25.83}{64.33}
          & \tfrecall{26.49}{51.15} & 6.36 \\
PronCFT-C & \tfrecall{\textbf{7.41}}{\textbf{72.14}}
          & \tfrecall{\textbf{10.33}}{\textbf{60.70}} & \textbf{5.68} \\

\bottomrule
\end{tabular}
\end{flowtable}

With correct phonemes, PronCFT recalls the replacement spelling
in 64.33\% of cases, compared with 6.11\% for PAC. When phonemes
are incorrect, PronCFT's original-spelling recall falls to 6.36\%,
whereas PAC retains 50.59\%. PronCFT-C further raises
replacement-spelling recall to 72.14\%, a 7.81-point improvement
over PronCFT ($p<0.001$). Together, the spelling and phoneme
interventions support the conclusion that PronCFT uses the content
of candidate pronunciations.

\subsection{Ablation study}
\label{ssec:expablation}

We test structured phoneme supervision and misspelling augmentation
(MS) using the four configurations in Table~\ref{tab:ablation}.
They share the same initial checkpoint, training utterances,
candidate phonemes, list construction, and training steps; only
structured supervision and MS vary. All are evaluated on identical
OOV contexts before curriculum compression.

\begin{flowtable}
\caption{OOV recall (\%) with phoneme-augmented contexts.
Struct. sup.: structured phoneme supervision;
MS: misspelling augmentation. ASR: transcription-only output;
Struct. out.: structured output.}
\label{tab:ablation}
\centering
\footnotesize
\setlength{\tabcolsep}{3pt}
\begin{tabular}{@{}lccrr@{}}
\toprule
Variant & Struct. sup. & MS & ASR & Struct. out. \\
\midrule
Phoneme-only   & No  & No  & 61.09 & --    \\
MS-only        & No  & Yes & 57.35 & --    \\
Structured-only & Yes & No  & 60.46 & 63.48 \\
PronCFT        & Yes & Yes & 65.45 & 74.12 \\
\bottomrule
\end{tabular}
\end{flowtable}

Under transcription-only decoding, neither component alone improves
on Phoneme-only (61.09\%): Structured-only reaches 60.46\% and
MS-only reaches 57.35\%. Combining both raises recall to 65.45\%.
With structured output, adding MS to structured supervision raises
recall from 63.48\% to 74.12\%. The two components work best together
in the tested configurations.

\subsection{Attention probing}
\label{ssec:expprobe}
Phoneme-annotated context improves OOV recall, but attention to
phonemes may reflect pronunciation matching (M1) or spillover from
adjacent spellings (M2). We probe this distinction on 1{,}000
utterances from the merged OOV test set by replacing only the target
candidate's phonemes, keeping spellings fixed, and measuring the
change in target--distractor phoneme attention ($\Delta P$;
Table~\ref{tab:probe}).

\begin{flowtable}
\caption{Attention probing on $n{=}1000$ utterances retained from the merged OOV test set. Following span-wise attention analysis~\cite{clark2019bert}, we teacher-force the reference transcript and pool attention at the target-hotword output positions (averaged over layers and heads) over each candidate's spelling and phoneme spans.}
\label{tab:probe}
\centering
\footnotesize
\setlength{\tabcolsep}{3.8pt}
\begin{tabular}{@{}lrrrrr@{}}
\toprule
Model & PhnR & $\mathrm{SAM}_t$ & $\Delta P$ & $\Delta P^{\mathrm{wph}}$ & PGD \\
\midrule
Qwen3ASR    & .212 & .0124 & .0022 & .0017 & $-21.8\%$ \\
Giga-SFT    & .233 & .0082 & .0015 & .0012 & $-17.4\%$ \\
Phn-SFT     & .113 & .0248 & .0023 & .0020 & $-13.2\%$ \\
\midrule
PronCFT-C   & .206 & .0280 & .0062 & .0035 & $-44.2\%$ \\
PronCFT     & \textbf{.346} & .0145 & .0063 & .0045 & $-29.0\%$ \\
\bottomrule
\end{tabular}
\end{flowtable}

PronCFT allocates a larger share of target-entry attention to
phonemes ($\mathrm{PhnR}=.346$) than Qwen3-ASR ($.212$),
Giga-SFT ($.233$), or Phn-SFT ($.113$), suggesting that exposure
to phoneme annotations alone does not explain the pattern.
PronCFT-C has a $\Delta P$ comparable to PronCFT ($.0062$
vs.\ $.0063$) but a lower PhnR ($.206$), alongside greater
attention to the spelling span ($\mathrm{SAM}_t=.0280$
vs.\ $.0145$).

Corrupting the target phonemes significantly reduces $\Delta P$
in every model: the relative reductions are 29.0\% and 44.2\%
for PronCFT and PronCFT-C, versus 21.8\%, 17.4\%, and 13.2\%
for Qwen3-ASR, Giga-SFT, and Phn-SFT. PronCFT-C's large
reduction despite its lower PhnR suggests that sensitivity to
phoneme content and preference for the phoneme span measure
different properties. However, its target log-probability rises
under corruption ($-10.32\to-9.68$), while changes in the other
models are below $0.11$ nat, so the attention shift alone does
not establish audio--phoneme matching. The probe therefore
characterizes attention allocation and sensitivity rather than
demonstrating an end-task gain.

\FlushFlowTables

\section{Conclusion}
\label{sec:concl}
PronCFT explicitly trains phoneme transcription, context-candidate alignment, and spelling generation, improving phoneme-aware OOV recall from 59.89\% to 83.57\% (+23.68\,pp). Future work will study how different ways of removing auxiliary outputs affect the model's ability to match candidate pronunciations to speech and its OOV recall.

\FlushFlowTables
\newpage
\bibliographystyle{IEEEbib}
\bibliography{refs}
\end{document}